\documentclass[twocolumn,showpacs,preprintnumbers,amsmath,amssymb]{revtex4}

\usepackage[dvips]{graphicx}
\usepackage{dcolumn}
\usepackage{bm}

\newcommand{\nc}{\newcommand}
\def\nn{\nonumber\\}
\def\bea{\begin{eqnarray}}
\def\eea{\end{eqnarray}}
\nc{\braket}[1]{\langle\,{#1}\rangle}
\def\wt{\widetilde}
\def\Z{\mathbf Z}  \def\R{\mathbf R}  

\def\Vec#1{\mbox{\boldmath $#1$}}

\begin{document}

\preprint{}

\title{2D Lattice Liquid Models}

\author{Yukitaka Ishimoto, Takahiro Murashima, Takashi Taniguchi, and Ryoichi Yamamoto}
\affiliation{Dept. of Chemical Engineering, Kyoto University, Japan}

\date{February 20, 2012}

\begin{abstract}
A family of models of liquid on a 2D lattice (2D lattice liquid models) have been proposed as primitive models of soft-material membrane. As a first step, we have formulated them as single-component, single-layered, classical particle systems on a two-dimensional surface with no explicit viscosity. Among the family of the models, we have shown and constructed two stochastic models, a vicious walk model and a flow model, on an isotropic regular lattice and on the rectangular honeycomb lattice of various sizes. 
In both cases, the dynamics is governed by the nature of the frustration of the particle movements.
By simulations, we have found the approximate functional form of the frustration probability, 
and peculiar anomalous diffusions in their time-averaged mean square displacements in the flow model. The relations to other existing statistical models and possible extensions of the models are also discussed.
\end{abstract}


\vspace{100pt}                              
\maketitle

\section{Introduction}

Two-dimensionally extended materials, or membranes, have been paid much attention to in various fields of study, such as lipid bilayers \cite{SakumaIT,Lipids,LipidSim} and surfactants \cite{Surfactant} in soft-matter and biophysics \cite{Various}. Alongside, many theoretical studies on those membrane-type objects have been done toward a general understanding of physical properties of such materials \cite{LipidSim,Simulation,FieldTheory}. Some of them were simulation approaches \cite{LipidSim,Simulation} and others were with field theoretical techniques \cite{SakumaIT, FieldTheory}. For example, one of the authors studied the pore formation in a binary giant vesicle of two-component lipid bilayer, and reproduced its dynamical shapes in the field-theoretical approach with some numerical calculations \cite{SakumaIT}. Although most of the works contributed successfully to our knowledge of individual membranes, it can be said that we are yet very far from the general understanding.

In this work, we aim to give some new insight to the general understanding of the physical properties of soft-material membranes by studying mathematical models on discretized surfaces both analytically and numerically. That is to say, we propose a family of fluid particle models of two-dimensional lattice as simpler models of membranes: tightly pack molecular particles on a given lattice, and give the particles' dynamics by some stochastic process, or by a set of mathematical rules. For more simplicity, we consider single-component, single-layered incompressible systems, and assume no explicit viscosity and an upper bound for the flow velocities on the particles. After the formulation, we construct two stochastic models of 2D lattice liquid, as our first steps. One of them is found to be the vicious random walk \cite{fisher} in its densest phase and a version of the dynamic lattice liquid model (DLL model) \cite{Pakula} in two dimensions. The dynamics of these models is given by completely independent time steps and would not generate various fluid dynamical phenomena such as lasting convective flows, while our formulation does not exclude such possibilities. In fact, the other model is constructed as the flow model with a mathematical rule, which has no corresponding known models. We further restrict ourselves to a finite flat honey-comb lattice and simulate the two models with their highest mobility. Comparing with similar lattice models such as various lattice random walks or the loop gas model (SOS model) \cite{Kostov89}, we show some statistical properties of the models by the simulation results.

The results tell that the flow model may exhibit some anomalous diffusion, in terms of time-averaged mean square displacements (TAMSD), and the ergodicity with its fairly long relaxation time. Relations to other models and some possible extensions of the models will also be discussed in the final section.

\section{Preliminaries}

There are various situations of the membrane systems, and each situation would deserve a single mathematical model in order to simulate it. However, since we aim to construct a basic model to be utilised for those various situations, we focus exclusively on a very simple but basic situation among them as a building block. 
First, we focus on a single-component and single-layered molecular system on a two-dimensional surface, treating the surrounding medium as a kind of heat bath to the surface. In other words, we introduce some rules or external forces onto the elements of the membrane rather than consider the interactions between the membrane and the medium.
Secondly, we focus exclusively on a simple classical interaction such as ``collision'', neglecting any other complicated interaction between the elements of the membrane. Let us call this `no explicit viscosity'.
Thirdly, we assume that the density fluctuation can be negligible and two elements would not occupy the same location of the surface at any time. This would correspond to the incompressibility condition of the surface from the two-dimensional fluid point of view.
To summarise, we assume and focus on a single-component and single-layered incompressible particle systems with no explicit viscosity on a two-dimensional surface exposed in an external (random) field.


Now, we are ready to define the 2D lattice liquid models.
Given a two-dimensional lattice $\Lambda$, a 2D lattice liquid model can be defined as a mathematical model of the particles which can only reside at the sites of the lattice $\Lambda$, covering all or almost all of the sites, and which flow inside $\Lambda$ as molecular-dynamical units of liquid do inside them.
Here, the particles represent the elements of the membrane, and are assumed to be of the same physical properties: their sizes, masses, and so on. 
The word `liquid' comes directly from two parts of the above definition, one of which is that the particles cover all or almost all of the sites of $\Lambda$, and the other of which is that the particles flow. 
Obviously, there is an ambiguity on the phrase ``all or almost all''. To avoid this ambiguity in this paper, we force each site to hold only and exactly one particle at a time, covering {\it all} the sites. We call it ``the occupation number is exactly one everywhere''. 
Although we can make some of the sites empty or put as many particles as possible, the assumption of the density fluctuation eliminates the latter and we simply omit the former in this paper. 
Note that the definition of the lattice liquid would not exclude these cases and would be worth studying elsewhere. 

One can consider the boundary conditions in the presence of boundary of the surface.
There are two types of boundary conditions, one of which is on particles' dynamical degrees of freedom, and the other of which is on their internal degrees of freedom, such as particle species. 
The former type of boundary conditions are naturally embedded in the lattice on which we define the model.
On the latter, since we only consider single-component systems, there is no boundary condition on the particle species. 

To complete the definition of the model, the particles' ``flows'' should be defined by imposing a set of mathematical rules on the flows. This consequently determines the dynamics of the system. Before going into the details of the rules, let us introduce the lattice conventions and consider the configuration space of the model to begin.

\subsection{The configuration space of the model}

A lattice $\Lambda$ is defined by a set of sites $i\in V(\Lambda)$ and a set of bonds $e_{i,j} \in E(\Lambda)$. A bond $e_{i,j}$ is connecting a pair of the sites, $i,j \in V(\Lambda)$, which are called neighbouring sites or in the neighbourhood. The lattice spacing $a$ -- the shortest length of the bonds specifies a fine aspect of the lattice. We consider only the cases with the same bond length, unless otherwise stated. For that reason we set $a=1$ without loss of generality.
The total number of the sites and the total number of the bonds are denoted by $dim(V(\Lambda))$ and $dim(E(\Lambda))$, respectively. We abbreviate them by $dim V$ and $dim E$ for simplicity.
The degree of the site, $deg(i)$, is the number of the bonds connected to the site $i$. If there is only one value for $deg(i)$ on $\Lambda$, we write it by $deg$ for short, and we consider mainly such cases.
In the later sections of some numerical calculations, we focus on the honeycomb lattices of various sizes as a series of examples. 

On a given lattice $\Lambda$, under any set of the rules of the movements, a flow of a particle can be sketched schematically in Fig.\ref{fig:worldlines} below.
\begin{figure}[h]
  \includegraphics[width=8cm]{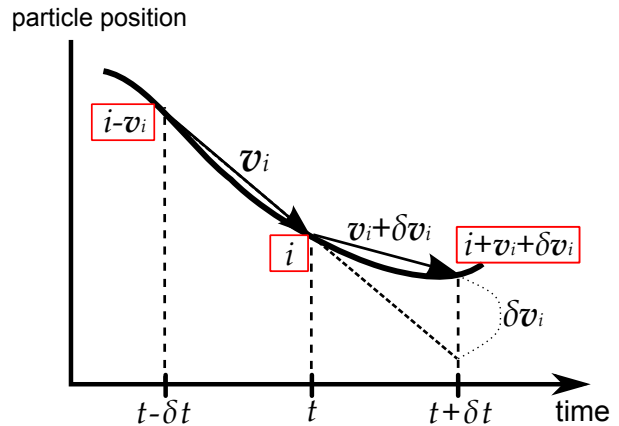}
\caption{One of the ``world-lines'' illustrating a particle travelling on a lattice.}
  \label{fig:worldlines}
\end{figure}
Introducing the minimum time lapse $\delta t >0$, a particle at the site $i$ at the time $t$ flew from the site $(i-\Vec{v}_i)$ at $t-\delta t$ with the velocity $\Vec{v}_i$, where $\delta t$ is implicitly taken as a unit ($\delta t=1$). Then, the particle at $i$ at $t$ flows toward $(i+\Vec{v}_i+\delta \Vec{v}_i)$ at the future time $t+\delta t$, subject to some fluctuation $\delta\Vec{v}_i$ by external forces. Explicitly writing the time dependence of the variables, $\Vec{v}_i(t)$ denotes the flow velocity of the particle from $t-\delta t$ to $t$, and $\Vec{v}_i(t)+\delta \Vec{v}_i(t)$ does from $t$ to $t+\delta t$. We denote the latter velocity by a redundant variable $\Vec{v}_i^\prime(t)\equiv\Vec{v}_i(t)+\delta \Vec{v}_i(t)$ for convenience.
The solid continuous line in Fig.\ref{fig:worldlines} represents the particle travelling in $(2+1)$-dimensional space-time, and all the particle movements can be represented by such world-lines. Namely, the configuration of the system at a certain time $t$, can be uniquely given by a set of the flow velocities: 
\bea
\{\Vec{v}_i(t)\} {\rm ~~for~all~} i \in V(\Lambda).
\eea
Accordingly, the time evolution is realised by mapping a configuration $\{ \Vec{v}_i(t) \}$ to another configuration $\{ \Vec{v}_i(t+\delta t) \}$ at a successive time, while a particle at $i$ travels to $i^\prime$ with the velocity $\Vec{v}^\prime_i(t)$ such that:
\bea
  i^\prime = i+\Vec{v}^\prime_i(t) , \quad 
  \Vec{v}_{i^\prime}(t+\delta t) = \Vec{v}^\prime_i(t).
\eea
For consistency, $\Vec{v}_i$ and $\Vec{v}_i^\prime$ cannot take completely arbitrary values, but are subject to the constraint that $\{\Vec{v}_i\}$ and $\{\Vec{v}_i^\prime\}$ map $V(\Lambda)$ to itself and they are bijective. In other words, the configuration space is given by the whole set of such bijections. Such a space is highly complicated in general, however one assumption actually makes the space more tractable.

We assume the following upper bound for the velocities:
\bea
  |\Vec{v}_i| \leq 1 {\rm ~~for~all~}i \in V(\Lambda).
\eea
This means that the particles are prevented from jumping in between non-nearest neighbouring sites. 
Therefore, the assumption is not too restrictive in the physics context.
When the bond lengths are all the same, this leads the absolute value $|\Vec{v}_i|$ takes either $0$ or $1$. Accordingly, the velocity $\Vec{v}_i$ takes one of $(deg(i)+1)$ values. Since $|\Vec{v}_i^\prime|$ obeys the same, the velocity change $\delta \Vec{v}_i$ takes at most one of $(deg(i)+1)$ values as well.

\begin{figure}[h]
\begin{minipage}{0.47\hsize}
  \includegraphics[width=3.5cm]{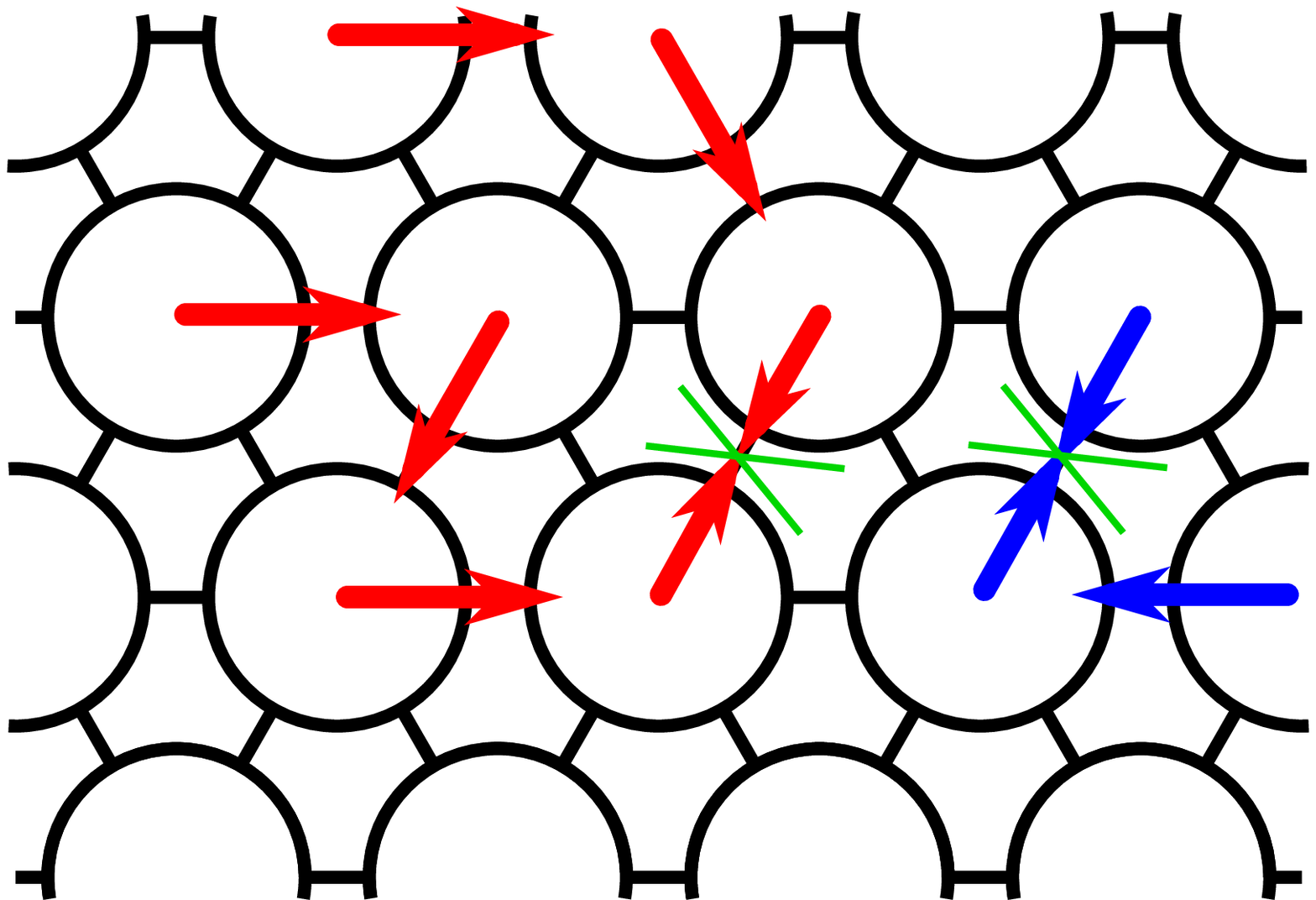}\\
  (a)
\end{minipage}\hspace{1em}
\begin{minipage}{0.47\hsize}
  \includegraphics[width=3.5cm]{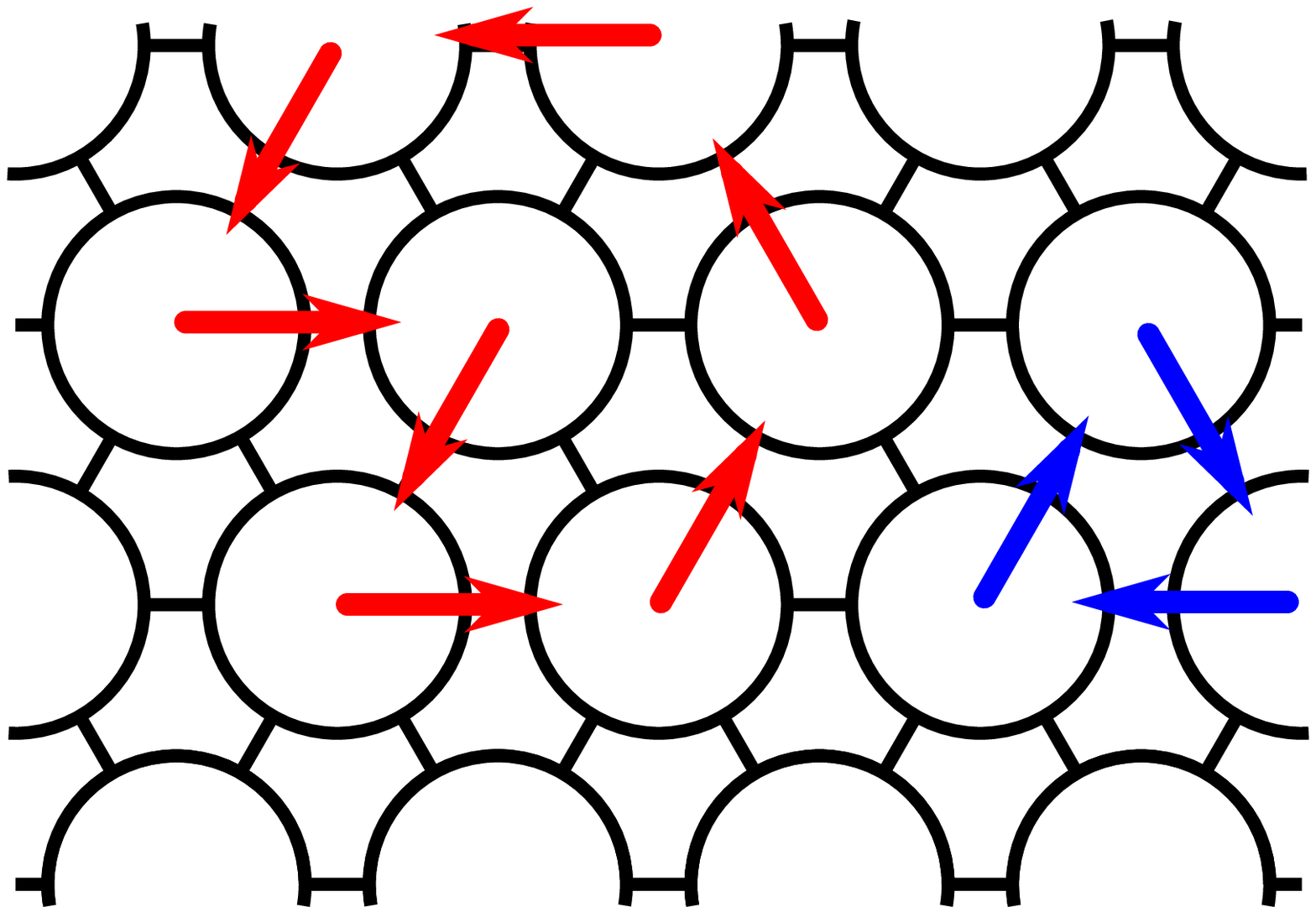}\\
  (b)
\end{minipage}
  \caption{(a) An example of the frustration of the velocities on a triangular lattice. Both red and blue sequences of vectors show the frustration. (b) An example of the possible movements on a triangular lattice. Red and blue sequences of vectors show the circulations.}
\label{fig:frustration-circ}
\end{figure}

With the upper bound of the velocities and the assumption of the occupation number, a configuration of the particles can be represented by a collection of oriented non-intersecting loops on $\Lambda$. In order to illustrate this fact, let us give a set of random velocities, $\delta \Vec{v}_i(t)$, with the condition $\Vec{v}_i(t)=0$. 
All particles potentially try to move toward the directions they have been kicked by some external forces. However, this does not necessarily give movements to the particles due to the frustration of such momenta (Fig.\ref{fig:frustration-circ}(a)). The frustration is an outcome of the excluded volume effect of the particles together with the restriction of the occupation number. 
In order to generate convective flows, some part of the ``potential'' movements generated by the forces must be enclosed so that the particles can move (Fig.\ref{fig:frustration-circ}(b)). 
Hence, the relevant part of the configuration $\{\Vec{v}_i\}$ can be represented by a collection of {\it oriented} non-intersecting loop graphs on $\Lambda$. We label the space of the oriented non-intersecting loop configurations by $\hat {\cal L}(\Lambda)$. We also label the loop configuration without the orientations by ${\cal L}(\Lambda)$ for later use.
Formal definition of the (oriented) loop configurations in graph theory, refer to Appendix \ref{ap:loops}.
Note that, if and only if $\Vec{v}_i(0)=0$, the above algorithm coincides with a version of the DLL model \cite{Pakula}. In what follows, we focus on the situations where the loop configuration solely plays a crucial role.

\subsection{Formulation of the model}

As was stated before, the set of mathematical rules of the particle-flows give the principal part of the definition of the model. However, regardless of how and what sort of the rules are imposed, they are to be expressed in the form of the velocities and the velocity changes: $\{\Vec{v}_i\}$ and $\{\delta \Vec{v}_i\}$, or in the form of the present loop configuration and the resulting one, $\hat G, \hat G^\prime \in \hat {\cal L}(\Lambda)$, representing the velocities. So, we formulate the 2D lattice liquid models as a family of statistical models defined with such variables.

One of the most straightforward expressions of the propagator would take the form of: 
\bea
  \label{eq:1stprop}
  &&
  {\cal G} ( x_1, x_2, \cdots ; \hat G, t ; \hat G^\prime, t^\prime)
  \nn&&
  {\rm or~~} 
  {\cal G} ( x_1, x_2, \cdots ; \{\Vec{v}_i(t)\}, t ; \{ \Vec{v}_j(t^\prime) \}, t^\prime), 
\eea
or as its ensemble: 
\bea
  \label{eq:2ndprop}
  &&
  Z^g( x_1, x_2, \cdots ; \hat G, t ; t^\prime) 
  \nn&&
  \equiv \sum_{\hat G^\prime \in \hat {\cal L}(\Lambda)} {\cal G} ( x_1, x_2, \cdots ; \hat G, t ; \hat G^\prime, t^\prime),
\eea
where the $x_k$'s are other unspecified parameters to prescribe the propagator.
If all the processes are statistical from the beginning, $i.e.$, even the initial configuration is given by some distribution, the following quantity would be the partition function in question:
\bea
&&   Z(x_1, x_2, \cdots ; t,t^\prime) 
   \nn&&
   \equiv \sum_{\hat G \in \hat {\cal L}(\Lambda)} P_{init} \left( \hat G \right) Z^g( x_1, x_2, \cdots ; \hat G, t ; t^\prime) , 
\eea
where the function $P_{init}(S)$ is the statistical weight of the configuration $S$. Practically, we will focus on the second expression (\ref{eq:2ndprop}) of the propagator and the first one (\ref{eq:1stprop}) for the construction of the second.

Let us construct a more concrete yet abstract form of the propagator by the probabilities of the configurations. 
For the time step of $\delta t$, given the present configuration $\hat G$, let us define $p(i)$ as the ``potential'' probability of a particle remaining at the site $i$, and $x(i;j)$ is the ``potential'' probability of the particle at $i$ being kicked toward one of the nearest neighbouring sites $j$. They satisfy: 
\bea
\label{eq:px-const}
  p(i)+\sum_{j} x(i;j) = 1,
\eea
where the summation is over all nearest neighbouring sites to $i$. 
In general, $p(i)$ and $x(i;j)$ may carry the information from the past, but we ignore it in the rest of this manuscript. Recall that some of the kicks with the ``potential'' probabilities might not be performed due to the before mentioned frustration. 
Suppose non-frustrated part of the particles are expressed in the configuration $\hat G^\prime$. The complementary graph can be given with its projection $G^\prime$ on ${\cal L}(\Lambda)$ by $\bar {\hat{G^\prime}} = \bar G^\prime$.
Then, the probability of having $\hat G^\prime$ in $\Lambda$, $P(\hat G^\prime)$, can be given by the product of the loop probability $P_l$ of having $\hat G^\prime$ in $G^\prime$ and the probability $P_f$ of the frustration in $\bar G^\prime$ as follows: 
\bea
  P(\hat G^\prime) = P_l(\hat G^\prime) \cdot P_f(\bar G^\prime),
\eea
where the initial configuration $\hat G$ is implicit, and the loop probability can be expressed by multiplying the probabilities $x(i;j)$'s along the oriented loops:
\bea
  P_l(\hat G^\prime) \equiv \prod_{\Vec{e}_{i,j} \in E(\hat G^\prime)} x(i;j).
\eea
Note that the definitions of the above probabilities need some minor extensions to the subgraphs for their consistency. For the detailed definition of the probability functions and their properties, refer to Appendix \ref{ap:PfPl}.

The probability $P(\hat G^\prime)$ is nothing but the first expression (\ref{eq:1stprop}) of the propagator for one time step.
Accordingly, the second expression (\ref{eq:2ndprop}) of the propagator for one time step is:
\bea
  Z^g(\hat G, t; t+\delta t) 
    &=& \sum_{\hat G^\prime \in \hat {\cal L}(\Lambda)} P( \hat G^\prime ).
\eea
Specifying its $\hat G$ dependence by $P(\hat G^\prime; \hat G)$, the propagator for a longer time span can be expressed by:
\bea
&&
  Z^g(\hat G^{(0)}, t; t^\prime)
\nn&& 
  = \!\!\!\!
  \sum_{\hat G^{(1)} \in \hat{\cal L}(\Lambda)} \cdots \sum_{\hat G^{(K-1)} \in \hat{\cal L}(\Lambda)} \prod_{k=1}^{K} 
   P(\hat G^{(k)}; \hat G^{(k-1)}),
\eea
where $t^\prime = t + K \delta t$. 
Hence, the construction of the propagator reduces to a detailed construction of the probability functions, $P_f$ and $P_l$.
Note that the above expression actually corresponds to the discretised version of the path integral.
Given the probabilities, an expectation value of some physical quantity $O$ for $\delta t$ can be given as usual by the form of: 
\bea
   \braket{O}_{\hat G} &=& \sum_{\hat G^\prime \in \hat {\cal L}(\Lambda)} O \cdot P(\hat G^\prime; \hat G). 
\eea

\section{Models}

It is obvious that the set of the mathematical rules of the particle-flows give the principal part of the definition of the 2D lattice liquid model as well as of its formulation. In fact, one can easily imagine the following three sorts of the rules for the particle-flows.
Namely, the particles flow according to some stochastic process, or subject to some external thermodynamic forces from the surrounding medium, or obeying some physics laws with their (fluid) dynamical properties such as the Langevin-type equation or the Navier-Stokes equation in the low $T$ single-component limit. 
In order to avoid ambiguities on discretisation and limiting procedures in the equation-based approaches, we deal only with the stochastic cases here. 
Some relations to thermodynamic or equation-based approaches will be discussed later.

\subsection{Stochastic Markovian model (I)}

In this section, we construct and formulate one of the simplest stochastic models, the stochastic Markovian model, whose rules are as simple as assigning simple kicks to push the particles and arresting them at the end of each step. The model is turned out to be equivalent to the vicious random walk in its densest phase and a version of the DLL model on two-dimensional lattice. The precise definition of the rules are given as follows.

First, let $p(i)$ be non-negative constant, and $x(i;j)$ be of isotropic kicks such that: 
\bea
  \label{eq:bounds4p}
  p(i) &=& p \quad {\rm for~0\leq p \leq 1},
  \nn
  x(i;j) &=& x(i) \equiv \frac{1-p}{deg(i)}.
\eea
Here, we restrict them to be strictly independent of the past, ensuring that the process is Markovian. 
Second, we assign the kicks according to the above probabilities, $p$ and $x(i)$, and those kicks are projected onto 
some oriented loop configuration $\hat G$. Perform the moves of $\hat G$.
Thirdly, as a key part of the rules, we force: 
\bea
  \label{def:sto}
  \Vec{v}_{j}(t+\delta t) = \delta \Vec{v}_i(t) \quad {\rm for~ all~} i,
\eea
where $j = i+\delta\Vec{v}_i(t)$.
This means that the configuration of the velocities is set at every step by the fluctuation $\left\{ \delta \Vec{v}_i \right\}$ only. Accordingly, one of the two sets, $\{\Vec{v}_i\}$ and $\{\delta\Vec{v}_i\}$, becomes redundant. 
Note that $|\delta\Vec{v}_i|$ inherits the upper bound from $|\Vec{v}_i|$: $|\delta\Vec{v}_i|\leq 1$. Practically, it follows $|\delta\Vec{v}_i|=0,1$.

The probability of an oriented loop configuration $\hat G$ on the sublattice $G$ can be written by:
\bea
  P_l (\hat G) = \prod_{\Vec{e}_{i,j}\in E(\hat G)} x(i).
\eea
If $deg(i)$'s are all the same as of regular lattices, the above probability can be simplified as:
\bea
  P_l (\hat G; x) = x^{l(\hat G)},
\eea
where we specify $x$ as an argument, $x=(1-p)/(deg)$, and $l(\hat G)$ denotes $dim E(\hat G) = \sum_i |\delta \Vec{v}_i|$ for $\{ \delta\Vec{v}_i \}$ of $\hat G$, {\it i.e.}, the total length of the loops in $\hat G$. 
Note that $x$ plays the role of the fugacity of the bonds in $\hat G$.
If we further define the probability of a non-oriented loop configuration $G$ on the sublattice $G$, summing up all the orientations of $G$, it follows that:
\bea
  P_l (G; x) = 2^{\alpha(G)} x^{l(G)}, 
\eea
where $l(G)=l(\hat G)$ by definition, $\alpha(G)$ counts the number of the loops in $G$, and the factor `2' indicates the directions of each oriented loop in $\hat G$.

Accordingly, the propagator, or the partition function, can be written down by:
\bea
  \label{eq:part_sto}
  {\cal G}^{sto} (x) = \sum_{ G \in {\cal L}(\Lambda)} P_l(G;x) P_f (\bar G ;x),
\eea
which is equivalent to unity by definition. 
$P_f(\Lambda; x)$ gives the probability of the no-loop configuration. 
The probabilities $P_f$ and $P_l$ take physical values, $|P_f|\leq 1$ and $|P_l|\leq 1$ for $0\leq p \leq 1$. 
It should be mentioned here that if $P_f(G)$ is set to be one for any $G$ then the partition function reduces to that of the non-oriented non-intersecting loop configurations on a lattice, which is known as the loop gas model \cite{Kostov89}: $Z^{loop}(n,x)$ with $n=2$. So, $P_f(G)$ plays a crucial role in our model.

As in the definition (\ref{def:sto}), 
the present configuration is always trivial so that the above expression (\ref{eq:part_sto}) is exact at any $t$. Therefore, the propagator can be expressed by:
\bea
  Z^{sto}(x; t; t+\delta t) = {\cal G}^{sto}(x). 
\eea
Hence, it can be summarised and simplified to the propagator of the time lapse $t$ as follows:
\bea
  Z^{sto}(x; t) = \left[ {\cal G}^{sto}(x) \right]^{t/\delta t}.
\eea
The expectation value of a physical quantity $O$ is given in a conventional manner:
\bea
  \braket{O}_t &=& \left[ \braket{O} \right]^{t/\delta t},
\nn
  \braket{O} &=& \sum_{ G \in {\cal L}(\Lambda)} O\cdot P_f (\bar G ;x) P_l(G;x).
\eea
Before giving our simulation results, 
let us show some simple relations and calculations on an isotropic regular lattice.

\subsubsection{On an isotropic regular lattice}

On an isotropic regular lattice, such as a square lattice with periodicities, the one-point function of the velocity can trivially be given by:
\bea
  \braket{ \Vec{v}_i } = 0 , 
\eea
by symmetry. Similarly, the expectation value of the absolute value of the velocity at $i$ can be given by:
\bea
  \braket{ |\Vec{v}_i| } = \frac{\braket{\sum_j |\Vec{v}_j|}}{dim V}
  = \frac{\bar l}{dim V}
\quad {\rm ~for~ all~} i, 
\eea
where $\bar l$ is the expectation value of the length of the loops: 
$\bar l \equiv \sum_{G\in{\cal L}(\Lambda)} l(G) P_f(\bar G; x) P_l(G; x)$.
Besides, the variance of the velocity becomes:
\bea
  \braket{ (\Delta \Vec{v}_i)^2 }
  = \braket{ |\Vec{v}_i|^2 } = \frac{\bar l}{dim V},
\eea
where the normalised lattice spacing is used.

With the above facts, one can state that each particle behaves like a random walker on the given lattice with the constant trapping ratio of $\wt{p} \equiv 1-\braket{|\Vec{v}_i|}$, where a particle moves to a certain direction with the probability $\wt{x} \equiv \braket{|\Vec{v}_i|}/deg$. Namely, for any $t_0$ for large $\Lambda$ or small $t$:
\bea
  \braket{\Vec{x}_I(t_0+t) - \Vec{x}_I(t_0)} &=& 0 , 
  \nn
  \braket{ \left( \Vec{x}_I(t_0+t) - \Vec{x}_I(t_0) \right)^2 }
   &=& \frac{\wt{x} \cdot deg}{\delta t} \; t,  
\eea
where $\Vec{x}_I(t)$ denotes the position of the $I$-th particle at the time $t$.
However, it should be noted that the system is not a collection of independent simple random walks, as our particles do not occupy the same site at the same time. In fact, the system is in the densest phase of the vicious random walks --- mutually avoiding $N$ particles' random walks on a lattice with $N$ sites \cite{fisher}. The difference between this system and a collection of the independent simple random walks can be clearly shown by the spacial correlation of the velocities. For example, the spacial correlation of a nearest pair of the system is given by: 
\bea
  \braket{|\Vec{v}_i| |\Vec{v}_j|} \geq \frac{2 \bar l}{dim V \cdot deg} 
  \quad {\rm for~} \braket{i,j},
\eea
whereas those of the simple random walks is given by a simple product of one-point functions: $<|\Vec{v}_i||\Vec{v}_j|>=<|\Vec{v}_i|><|\Vec{v}_j|>= \left( \frac{\bar l}{dim V} \right)^2$.

\subsection{Stochastic flow model (II)}

In the previous sections, we have formulated the model where the streamlines are temporally generated and live only for the duration of $\delta t$. Because the probability of the shortest loop in a honeycomb lattice is at most $3^{-6}$ in the bulk, $\bar l$ seems to be very small.
We infer in the model (I) that 
the stochastic process realises the solid-like nature of the membrane rather than its liquid nature as the DLL model targeted supercooled liquids. In this section, we intend to explore another class of stochastic models where the particles keep flowing on the membrane rather than obeying the `stop and go' evolution.
In order to realise this phenomenon, we take the simplest way:
if no fluctuations are added to the system, the particles keep flowing along the oriented loops. This rule can be expressed by the relation in the absence of the fluctuations:
\bea
  \label{def:stoF}
     \Vec{v}_i(t+\delta t) = \Vec{v}_i(t).
\eea
Now, it is evident that the key point of the construction lies on how to add stochastically generated fluctuations onto the existing streamlines.  

In order to avoid complications for realising the relation (\ref{def:stoF}) and the frustrations in terms of $\{\delta \Vec{v}_i\}$, let us formulate the above addition in terms of the oriented loop configurations. 
Say, $\hat G_i$ is the initial configuration and 
$\hat G_f$ is the resulting configuration. 
The arithmetic rule (addition/subtraction) for the oriented loop configurations can be illustrated in a multiplicative form by: 
\bea
  \delta \hat G \cdot \hat G_i  = \hat G_f \in \hat {\cal L}(\Lambda),
\eea
where $\delta \hat G$ stands for the stochastically generated fluctuations, and vanishing $\delta \hat G$ is given by the no-loop configuration $I(\Lambda)$ naturally satisfying the rule (\ref{def:stoF}). In ${\cal L}(\Lambda)$, there is a natural definition of such an additive operation as the symmetric difference.
We extend and modify it to our case for $\hat{\cal L}(\Lambda)$. 

Let $\hat G_1$, $\hat G_2$ be elements in $\hat {\cal L}(\Lambda)$. 
If $\hat G_2$ consists of the same sites and bonds with opposite directions as those of $\hat G_1$, then we call $\hat G_2$ the inverse of $\hat G_1$, denoting it by $\hat G_2 = \hat G_1^{-1}$. The additive operation between $\hat G_1$ and $\hat G_2$ is defined in the multiplicative form by:
\bea
  \hat G_1 \cdot \hat G_2 = \hat G , 
\eea
such that the directed bonds of the elements satisfy: 
\bea
  \label{eq:symdiff-directed}
  \hat G &=& \hat G_1 \cup \hat G_2 - \hat G_1 \cap \hat G_2^{-1} 
     - \hat G_1^{-1} \cap \hat G_2 , 
\eea
while the disconnected sites are removed from $\hat G$. The above addition trivially satisfies the followings:
\bea
  \hat G_1 \cdot I(\Lambda) &=& I(\Lambda) \cdot \hat G_1 = \hat G_1 
      \quad {\rm for~all~} \hat G_1\in \hat {\cal L}(\Lambda),
  \nn
  \hat G_1 \cdot \hat G_2 &=& I(\Lambda) 
      \quad {\rm only~when~}\hat G_2 = \hat G_1^{-1},
\eea
where $I(\Lambda)$ behaves as the zero in the addition and the identity in its multiplicative form. The problem of this definition is that $\hat G$ is in the corresponding (directed) cycle space but not necessarily in $\hat {\cal L}(\Lambda)$. 
To avoid this unwanted situation, 
we propose the following projected addition for $\hat G$ in eq. (\ref{eq:symdiff-directed}):
\bea
 \label{def:addition}
 \hat G_1 \cdot \hat G_2 &=& \hat G
   \quad {\rm for~} \hat G \in \hat {\cal L}(\Lambda),
 \nn
 \hat G_1 \cdot \hat G_2 &=& \hat G_2
   \quad {\rm for~} \hat G \not\in \hat {\cal L}(\Lambda).
\eea
In the second line, we call $\hat G_1$ is non-additive to $\hat G_2$, or vice versa. This projection drastically changes the algebraic nature of the operation. In fact, it breaks the commutativity: $\hat G_1 \cdot \hat G_2 \ne \hat G_2 \cdot \hat G_1$, and the associativity: $(\hat G_1 \cdot \hat G_2) \cdot \hat G_3 \ne \hat G_1 \cdot (\hat G_2 \cdot \hat G_3)$. So, the ordering of the operations matters in this case. It should be mentioned here that we can create some different definitions of the addition simply by different projections of $\hat G \not \in \hat{\cal L}(\Lambda)$ onto $\hat {\cal L}(\Lambda)$. As a starting point, we stick to the above definition and will mention another possibility later. 
Note also that, precisely speaking, $\delta \hat G$ is either a candidate for a fluctuation or a fluctuation which sometimes does nothing, because of the projection.

If the configuration at the time $t_0$ is given by $\hat G(t_0)$, the configuration $\hat G(t_0 + t)$ at the time $(t_0 + t)$ can be expressed by a series of successive additions of $\delta \hat G$ to $\hat G(t_0)$:
\bea
   \hat G(t_0+t) = \left[ \prod_{k=0}^{t/\delta t -1} \delta \hat G(t_0+k \delta t) \right] \cdot \hat G(t_0).
   \label{def:flow-part}
\eea
The multiplication in the r.h.s. means the multiple application of the loop additions, $\prod_{i=1}^k A_i \equiv A_k \cdots A_2 \cdot A_1$, and the addition should be done from right to left. The factor expresses nothing but a configurational and kinetic path between two configurations.

Finally, we complete the definition of the stochastic flow model by stochastically giving $\delta \hat G$ in the same manner as in the previous section. Namely, assign the ``potential'' probabilities of a particle remaining at the same position and of a particle at $i$ being kicked toward one of the nearest site $j$ by $p$ and $x(i;j)$, respectively. For the isotropic distribution, $x(i;j)$ can be simplified to $x(i)=(1-p)/deg(i)$. Then, project the velocity changes onto the space of the oriented loop configurations $\hat {\cal L}(\Lambda)$, and add it to the current configuration in the end. The degrees of freedom of $\delta \Vec{v}_i$ and $\Vec{v}_i$ are same as $(deg(i)+1)$, as long as an additional upper bound for $\delta \Vec{v}_i$ is not imposed. So, their spaces differ only by the meanings of $p$ and $x(i;j)$. 
For simplicity, we neglect this difference and distribute the probabilities isotropically. 
Note that 
the addition rule implicitly introduced the upper bound for $\delta\Vec{v}_i$ such that $|\delta\Vec{v}_i|<2$.

Now, we have formulated how the fluctuations are given and added to existing flows. The propagator with the initial configuration $\hat G_i(t_0)$ can now be written down by:
\bea
&&\!\!\!
 Z^{flow}(x; \hat G_i,t_0; t+t_0) 
\nn&&\!\!\!
=  \!\!\!\!\!
   \sum_{\delta \hat G_{t/\delta t} \in \hat{\cal L}(\Lambda)} \! \cdots \! 
   \sum_{\delta \hat G_1 \in \hat{\cal L}(\Lambda)}  
   \prod_{j=1}^{t/\delta t} \left[ 
     P(\delta \hat G_j) \delta \hat G_j 
   \right] \cdot \hat G_i(t_0), \nn
\eea
where the probability $P(\hat G)$ of having the loop $\hat G$ in $\Lambda$ is the same as in the stochastic Markovian model.
Note the process of this model is also Markovian.

\section{Simulation results}

\begin{figure}[h]
\centering
  \includegraphics[width=5cm]{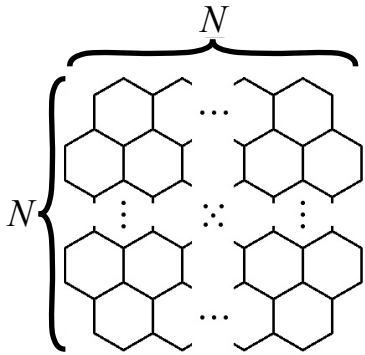}
\caption{The illustration of the rectangular honeycomb lattice $\Lambda_{h_N}$.}
  \label{fig:hN}
\end{figure}

In this section, we perform numerical calculations on the proposed models 
and show their statistical properties for further investigation, some of which reveal very interesting features of the models.
We restrict the lattice $\Lambda$ to be as simple as the finite rectangular honeycomb lattice of roughly $N\times N$ hexagons for odd $N$, labelling it by $\Lambda_{h_N}$. Precisely speaking, we define the lattice $\Lambda_{h_N}$ by piling $(N-1)$ and $N$ aligned hexagons alternatively up to $N$ rows so that the lattice becomes left-right symmetric (Fig.\ref{fig:hN}). 
For odd $N$, the lattice becomes actually top-bottom symmetric as well, and therefore symmetric under discrete rotations. 
In the case of a finite lattice with boundary, there are the boundary sites and bonds which face the outer region of the lattice on its planar embedding. Among such sites, some have smaller degrees from those in the bulk. In the case of the honeycomb lattice, there are degree-two boundary sites. Accordingly, we can assign two different probabilities: $x=(1-p)/3$ in the bulk and $x_b=(1-p)/2$ for the boundary sites of degree two. 
We choose the highest mobility $p=0$, $x=1/3$, and $x_b=1/2$ for numerical calculations. 

As two models share the oriented non-intersecting loop configurations as an input to the system, the input data has been generated in common for a million of time steps for various sizes, $N=11, 21, 31, 41, 51, 61$. Below is the basic information of $\Lambda_{h_N}$:
\bea
  dim V(\Lambda_{h_N}) &=& (N+1)(2N+1)-4,
\nn
  dim V_b (\Lambda_{h_N}) &=& 
  4N,
\nn
  dim E(\Lambda_{h_N}) &=& 
   3N^2 + \frac52 N - \frac92 ,
  \nn
  F &=& N^2 - \frac{N+1}{2} , 
\eea
for odd $N>1$. $V_b$ is the set of all the boundary sites of degree two. The number of faces is denoted by $F$.

\subsection{The statistics of the input data and the stochastic Markovian model}

\begin{figure}[h]
  \includegraphics[width=8cm]{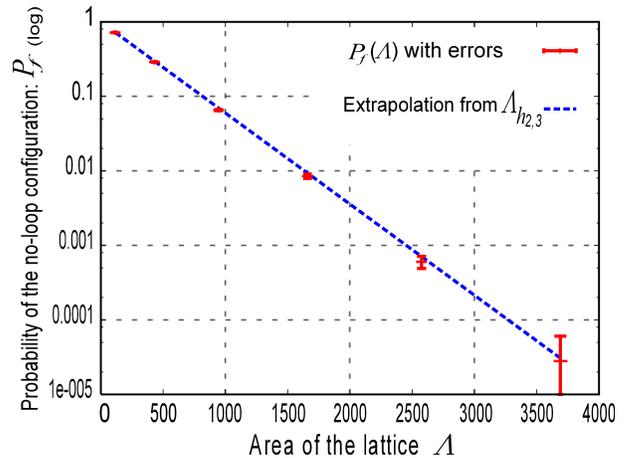}
\caption{The probability of the no-loop configuration for various sizes in the log scale in the $y$-direction. The $x$-axis stands for the area of the lattice counted by the number of hexagonal faces in $\Lambda$. There can be obviously seen a scalability. The error bars are given at the 95\% 
confidence level.}
  \label{fig:Pf-log}
\end{figure}

We first plot the probability $P_f(\Lambda_{h_N})$ of the no-loop configuration, or of the total frustration, for different sizes in Fig.(\ref{fig:Pf-log}). We have observed that it decays approximately exponentially with respect to the area of the lattice. 
Here,
we propose 
the approximate functional form of $P_f$ for general shape of the honeycomb lattice with $F$ connected faces:
\bea
  P_f(\Lambda) \simeq e^{- \frac{F}{n_0}}.
  \label{eq:Pfapprox}
\eea
Note the form may not well approximate if $\Lambda$ takes an extreme shape such as a lattice with small $F$ or of aligned faces. Fitting the data in Fig.\ref{fig:Pf-log} with the above function actually leads to $n_0\simeq 347.129 \pm 10.53$.
We have also calculated $P_f$ by hand in the case of the honeycomb lattice of 2 piles of 3 hexagons, $\Lambda_{h_{2,3}}$, and confirmed that the extrapolation from $\Lambda_{h_{2,3}}$ gives $ n_0 \simeq 355.089$. 
This result further indicates that the system gradually changes its nature between frustration dominant surface to loop dominant surface as the size changes around $N\sim 19$.

\begin{figure}[h]
  \includegraphics[width=8cm]{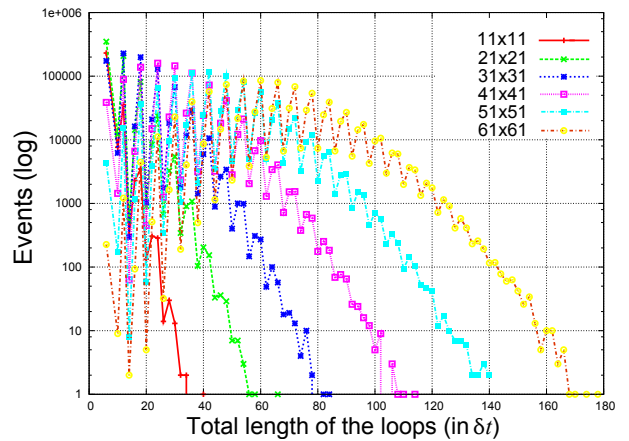}
\caption{ The distribution of the total length of the loops generated in one time step. The $x$-axis is the length in the unit of the lattice spacing. The $y$ shows the number of events in a million trials. Periodic oscillations seem to suggest the multiple entries of the same type of loops, e.g., two length-six loops at once contributes to the length-twelve event.}
  \label{fig:looplen_dist}
\end{figure}

Next, we plot the number of events for the total length of the loops $l(G)$, which is equivalent to the number of the moving particles in one time step (Fig.\ref{fig:looplen_dist}).
It is remarkable that the leading contribution is not necessarily from length-six loops, and neither is the sub-leading contribution from length-ten loops. For instance, on $\Lambda_{h_{41}}$, the leading and sub-leading contributions are from length-24 and length-30 loops, resp.
We have confirmed that the proposed form of $P_f$ can explain nicely the $N$-dependence of the probability of length-six loops.
We also report the proportionality of the mean loop length $\bar l$ w.r.t. the area of the lattice:
\bea
  \bar l \sim 0.01763 F, 
\eea
for $F\gg 1$.

As for the loop number distribution, 
we find that they obey the Poisson distribution in an accurate way with the mean $\bar \alpha$ approximately proportional to the number of the faces $F$:
\bea
  \bar \alpha \sim 0.002856 F.
\eea
The relation of the mean with the area of the lattice is not so trivial, suggesting that there is a relatively stable value of the loop number per unit surface as well as that of the total length of the loops per unit surface. 
Together with the average absolute value of the velocity and other basic numbers of $\Lambda_{h_N}$, we list them in Table \ref{table:hN}.
\begin{table}[h]
\begin{center}
\begin{tabular}{|r|r|r|r|c|c|c|c|}
\hline
$N$ & $dim V$ & $dim E$ & $F$ & $\bar l$ & $\bar \alpha$ & $\bar{|\Vec{v}|}$ \\\hline
11  & 272    & 386   & 115  & 2.01001   & 0.326687 & $7.38974 \times 10^{-3}$ \\
21  & 942    & 1371  & 430  & 7.558724  & 1.226437 & $8.02412 \times 10^{-3}$ \\
31  & 2012   & 2956  & 945  & 16.625344 & 2.695635 & $8.26309 \times 10^{-3}$ \\
41  & 3482   & 5141  & 1660 & 29.236702 & 4.737545 & $8.39653 \times 10^{-3}$ \\
51  & 5352   & 7926  & 2575 & 45.336774 & 7.346373 & $8.47100 \times 10^{-3}$ \\
61  & 7622   & 11311 & 3690 & 65.068168 & 10.540758 & $8.53689 \times 10^{-3}$ \\
\hline
\end{tabular}
\caption{ The basic numbers of $\Lambda_{h_N}$ for various $N$. }
\label{table:hN}
\end{center}
\end{table}
The very low rate of $\bar{|\Vec{v}|}$ means that each element of the membrane behaves like a random walk with a high ratio of trapping: $1-\bar{|\Vec{v}|}$, and the system exhibits a very slow dynamics compared to the unit of time, $\delta t$. 

\begin{figure}[h]
\centering
  \includegraphics[width=8cm]{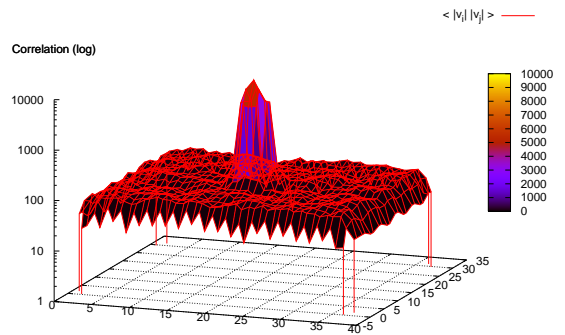}
\caption{An example of two-point correlation function of $\braket{|\Vec{v}_i||\Vec{v}_j|}$, one of which is fixed at the centre of the lattice $\Lambda_{h_{21}}$. The $z$-axis is the number of events in a million trials in the log scale.}
  \label{fig:2pt-dvdv452}
\end{figure}

In fact, we have confirmed that the model (I) shares the same property as the random walk in a cage in terms of the time-averaged mean square displacements (TAMSD):
\bea
\braket{ \Vec{x}_I(t)^2 }_{TA}
  &\!\!=& \!\! \frac{1}{t_e - t_i} \int_{t_i}^{t_e} \!\!\!\!dt^\prime  \left| \Vec{x}_I(t^\prime + t) - \Vec{x}_I(t^\prime) \right|^2 \!,
\eea
where we have used the discrete version of the above with $t_i=0$ and $t_e=10^6$ in the unit of $\delta t$.
To show the difference from the independent random walks, we further obtain the time-average of the spacial two-point correlation functions, $\braket{|\Vec{v}_i||\Vec{v}_j|}$, with $i$ fixed at the centre of $\Lambda_{h_{21}}$ (Fig.\ref{fig:2pt-dvdv452}):
\bea
  \label{eq:TAMSD}
  \braket{|\Vec{v}_i||\Vec{v}_j|} = \frac{1}{t_e-t_i} \int_{t_i}^{t_e}
     dt^\prime \braket{|\Vec{v}_i(t)||\Vec{v}_j(t)|}.
\eea
As you see, our model has the notable two-point correlation functions which are clearly distinguishable from those of the simple random walks.

\subsection{The statistics of the stochastic flow model}

\begin{figure}[h]
  \includegraphics[width=7cm]{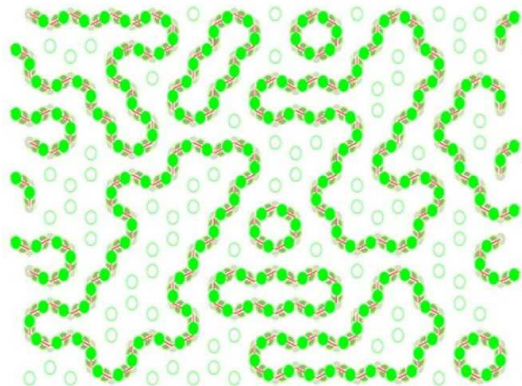}
\caption{An illustration of the lattice liquid on a honeycomb lattice. The sequences of green balls show the movements of the particles. The lattice is periodic to the horizontal direction.}
  \label{fig:animation}
\end{figure}

\begin{figure}[h]
  \includegraphics[width=8cm]{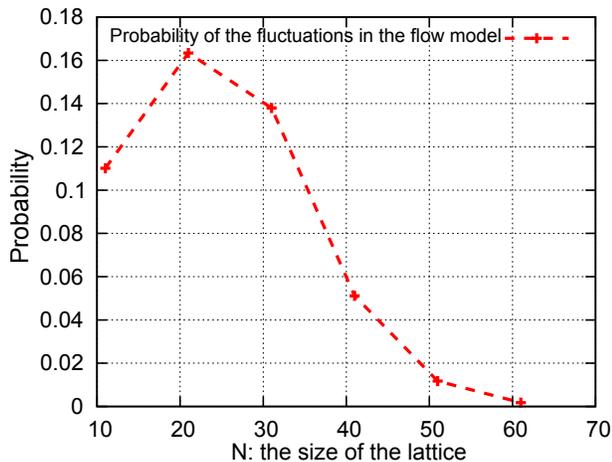}
\caption{The average probability of the fluctuations in the million steps. The probability suddenly drops down in between $N=31$ and $N=41$.}
  \label{fig:Pfluctuation}
\end{figure}

In the numerical calculation of the flow model, the initial configuration at $t=0$ is set to be the no-loop configuration $I(\Lambda)$ for all sizes.
The system gradually evolves as time goes, and finally reaches a quasi-critical state where all observable quantities become relatively stabilised. 
Their movements are schematically illustrated in Fig.\ref{fig:animation} in the case of the periodic $\Lambda_{h_{11}}$.
Although we have chosen the particular initial configuration, the data for various statistics is basically taken a certain amount of time after $t=0$ except for the probability of the fluctuations.
We count how many times the fluctuations are actually added to the system from the beginning to the end. Dividing it by the number of the input trials, we plot them in Fig.\ref{fig:Pfluctuation} as the probability.

For the size $N\geq 41$, the system becomes very stubborn about its stability against the ``potential'' fluctuations. Particularly for $N\geq 51$, much fewer fluctuations are added in contrast with the high probability of generating the ``potential'' fluctuations: $P_l(\Lambda_{h_N}) \sim 1$. Therefore, we will not look into these cases in detail, where the configurations are expected to be nearly frozen after it reaches a quasi-critical state. Note that there is no `critical state' such that no fluctuation can disturb, because there always exists at least one fluctuation trivially additive to the present configuration $\hat G$. It is the inverse $\hat G^{-1}$ that transforms it to $I(\Lambda)$ by the addition.

\begin{figure}[h]
  \includegraphics[width=8cm]{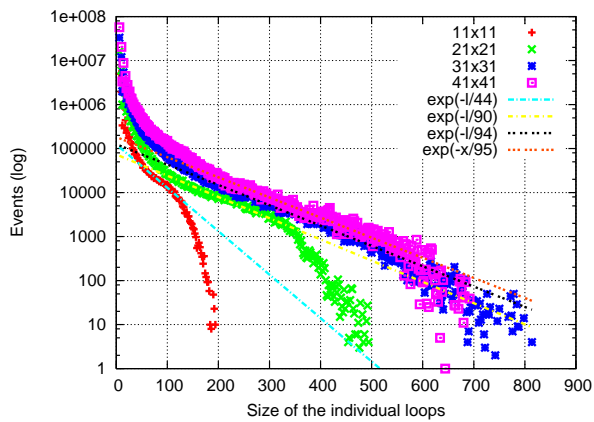}
\caption{The distribution of the size of the individual loops. There is an almost uniformly scaling region in the middle.}
  \label{fig:looplen-flow}
\end{figure}

Instead of plotting the total length of the loops, which is well stabilised in a quasi-critical state, we plot the distribution of the size of the individual loops in Fig.\ref{fig:looplen-flow}.
At each time step, we classify the loops by their perimeter lengths, and accumulate them w.r.t. the length. So, for example, if there is only one oriented one-loop of length six existing during a thousand of time steps, the number of events counted for the length-six loops is a thousand. The distribution is drawn as such for various sizes, and we have found neither sharp peaks nor obvious cut-offs for the loop sizes in the plot. The most probable loop is the smallest loop of length six, and it seems to play a leading role of the dynamics of the flow model.
There is no scaling relation in the entire region though, there is an almost uniformly scaling region in the middle part of the distribution.
Moreover, the slope of the scaling seems to change suddenly between $N=11$ and $N=21$. Therefore, we infer that the rise of the loop probability $P_l(\Lambda)$ for $\delta \hat G$ may affect the evolution and the size of the individual loops.

The average loop size, 
the average length of the loops $\bar l$, the average number of simultaneous loops $\bar \alpha$, and the average absolute value of the velocity $\bar{|\Vec{v}|}$ are all shown in Table.\ref{table:hN-flow}.
\begin{table}[h]
\begin{center}
\begin{tabular}{|c|c|c|c|c|c|}
\hline
$N$ & Ave. loop size & $\bar l$ & $\bar \alpha$ & $\bar{|\Vec{v}|}$ \\\hline
11  & 17.015 & 194.85 & 11.452 & 0.71638 \\
21  & 19.259 & 690.72 & 35.865 & 0.73325 \\
31  & 20.120 & 1484.8 & 73.797 & 0.73798 \\
41  & 20.327 & 2587.2 & 127.28 & 0.74303 \\
\hline
\end{tabular}
\caption{ The basic data of the stochastic flow model on $\Lambda_{h_N}$ for $N=11,\cdots,41$. }
\label{table:hN-flow}
\end{center}
\end{table}
The above averages include the data from the beginning, and therefore we neglected the relaxation time $\tau_N$ from $I(\Lambda)$ to quasi-critical states. The relaxation time $\tau_N$ which will be shown shortly is not very long compared to $t_{all} = 10^{6} \delta t$. So, it would not spoil the accuracy of the statistics much.
In this flow model, the distribution of the loop number behaves rather Gaussian than Poisson.
In addition, $\bar l$ and $\bar \alpha$ no longer obey simple proportional relations to the area of the lattice. The former has a rather correlation to the number of the sites $dim V$, as $\bar{|\Vec{v}|}=\bar l/dim V$ is almost independent of the size $N$.
The value $\bar{|\Vec{v}|}$ gives the ratio of the moving particles among all, while $1-\bar{|\Vec{v}|}$ is of the trapped particles. Nearly constant value of $\bar{|\Vec{v}|}$ might suggest that the value is intrinsically universal for all honeycomb lattices.
The average loop number $\bar \alpha$ is not so stabilised compared to the average loop length $\bar l$. It should be remarked here that the projection in our addition (\ref{def:addition}) would certainly change the probability of the fluctuations, but only its characteristic time scale. So, we claim that the above values and qualitative arguments are characteristic in this type of models, regardless of the addition rule defined earlier.

\begin{figure}[h]
  \includegraphics[width=8cm]{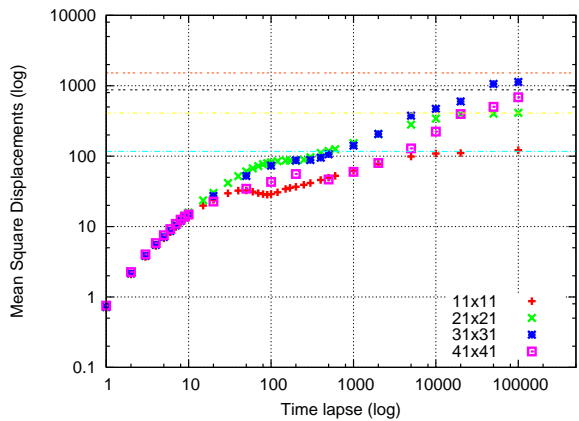}
\caption{The time-averaged mean square displacements (TAMSD) of the stochastic flow model (II) on $\Lambda_{h_N}$, for various sizes in the log-log format. The $x$-axis is the time lapse $t$, while the $y$-axis is the square distance in the unit of the lattice spacing squared. The dashed lines are all exactly calculated asymptotic values for various $N$.}
  \label{fig:MSD11-41}
\end{figure}

It is the dynamical aspect of the model that the addition rule makes different.
In this stochastic flow model, the particles are basically trapped in the sites alike in the stochastic Markovian model. But, once they are involved in the streamlines, they keep moving until it is stopped by a fluctuation. 
Coarse-graining of the pathway actually make itself look like a Brownian motion, but it is actually much more complicated. Let us quantify it by the TAMSD of eq. (\ref{eq:TAMSD}).
We picked up a particle situated at the centre of the lattice and calculated the TAMSD (Fig.\ref{fig:MSD11-41}), where the lower limit of the time is set by the relaxation time $t_i=\tau_N$: $\tau_{11,21}\sim 10^4 \, \delta t$, $\tau_{31,41}\sim 10^5 \, \delta t$, so that at $t\simeq \tau_N$ the particles are randomised enough on $\Lambda$ from the initial configuration. $t_e$ is set to be the end of the simulation: the million-th step.
As one can see from the left side of Fig.\ref{fig:MSD11-41}, the particle starts moving at a certain rate of $\bar{|\Vec{v}|}$, and diffuses like liquid, slightly faster than the Brownian motion. At around $t\sim 10$, the exponent $\nu$ in $\braket{\Vec{x}(t)^2}\sim t^\nu$ reaches the Brownian one, and later decreases to the anomalous diffusion of $\nu<1$ in a strange way. Finally, it approaches its asymptotic value. By the times the particles approach the asymptotic values, we can read off the relaxation times, $\tau_N$. Their orders of magnitude are roughly the same we set for the initial time $t_i$ for TAMSDs.
It can be said that the system is generally in an anomalous diffusion process where $\nu\geq 1$ in the short range, and $\nu<1$ in the distance. There can be found a strange and interesting plateau as well in the middle of the curve for each size of the lattice, but we have not yet obtained any clear explanation of the plateau. We will further mention this in the final section.

There is a closest model to compare with our model (II), which is the non-reversal random walk. In that model, a random walker can bring a little memory for its past path so that it will not come back to the memorised positions. Usually, the walker memorises only the previous position in order not to come back at once. The model is sometimes known as a polymer chain model with excluded volume. In our flow model, a particle movement is always associated with a ring of movements and cannot reverse immediately. This minimum regulation matches with that of the non-reversal walker. In addition, because the non-reversal walker is not bounded by any loop-like relation to neighbours, we deduce that it gives the upper limit for the TAMSD of our model. 

\begin{figure}[h]
\centering
  \includegraphics[width=8cm]{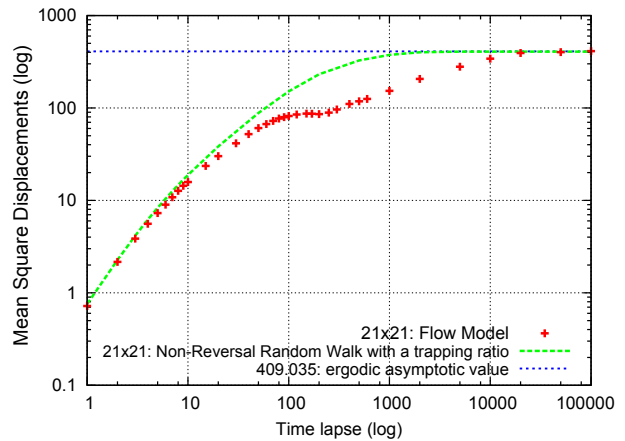}
\caption{A comparison of TAMSD between different models with flows on $\Lambda_{h_{21}}$. The line for the non-reversal random walk is given with the trapping time consistent with the original model, and the asymptotic value for $\Lambda_{h_{21}}$ is shown.}
  \label{fig:comparisonII_21}
\end{figure}

We define the stochastic flow model (II) and the non-reversal random walk on the lattice $\Lambda_{h_{21}}$, and compare them in terms of their TAMSDs in Fig.\ref{fig:comparisonII_21}.
Notice that the constant trapping ratio is equipped with the non-reversal walkers to keep the pace of his first step with the flows in our model.

As for the ergodicity, we claim that the ergodicity weakly holds in our model (II) of the lattice liquid. In order to illustrate this, we have calculated the asymptotic value of the MSD with some assumption as follows.
Assume that, in the limit of $t\to +\infty$, the time correlation of every particle on $\Lambda$ vanishes so that a particle situated at any of the sites on $\Lambda$ at a certain time $t_0$ is to be drifted to every site with equal footing in the infinite future. Then, the asymptotic value of $\braket{\Vec{x}_I(t)^2}$ can simply be given by the average of the square distances of all the relevant pairs of the sites on $\Lambda$. 
On the rectangular honeycomb lattice $\Lambda_{h_N}$, we derived its exact formula and the leading terms are given by:
\bea
  \label{eq:asymptotic_approx}
  \braket{\Vec{x}_\infty^2} \equiv \lim_{t\to +\infty} \braket{\Vec{x}_I(t)^2} 
  = \frac18 \left( 7 N^2 + 10 N \right) + O(1).
\eea
In the case of $N=11,21,31,41,51$, the exact values are up to first five digits:
\bea
  \label{eq:asymptotic}
  \braket{\Vec{x}_\infty^2} \simeq 116.75, 409.04, 876.45, 1518.9, 2336.4.
\eea
See Appendix \ref{app:aymptotic} for the exact form of an arbitrary $N$.

The time-averaged MSDs in Fig.\ref{fig:MSD11-41}, \ref{fig:comparisonII_21} approach the calculated asymptotic values (\ref{eq:asymptotic}) and are stabilised around them as the time grows. Therefore, we can state that our models are surely ergodic in this sense.

\section{Concluding remarks}

In this paper, 
we have proposed and formulated a family of models, called the 2D lattice liquid models, as one sort of the simplest models of single-component, single-layered, incompressible fluid particle system on a membrane without explicit viscosity and density fluctuation. The model is formulated as a particle system on a lattice, but the configurations are found to be expressed by the sets of oriented non-intersecting loops, due to the restriction on the occupation number and the upper bound for the velocities. 
Following the formulation, we have constructed two models of the lattice liquid. One of them is the stochastic Markovian model (I) which is found to be another manifestation of the vicious random walks in its densest phase and the DLL model in two dimensions. We have dealt with the model on the isotropic lattice, and derived a few basic quantities of the model. We have also shown that the difference between our model and a collection of non-interacting simple random walks by distinct two-point functions.
The other model is the stochastic flow model (II) where the set of rules are given in eqs.(\ref{def:stoF}, \ref{eq:symdiff-directed}, \ref{def:addition}). The first rule (\ref{def:stoF}) is on the stability of the existing flow and is written in the Euler formulation. We have utilised the notion of the oriented loop configurations $\hat {\cal L}(\Lambda)$ and formulated the model.
 In order to interpret the fluctuation as a direct consequence of the kicks by the environment, it is worth rewriting the rule in the original Lagrangian formulation. In this paper, we have focused on the construction of the model itself, and neglected this point.

In numerical calculations, we have realised both models on the rectangular honeycomb lattice $\Lambda_{h_N}$ of various sizes in one million time steps. 
From the input data of the models on $\Lambda_{h_N}$, we have found that the probability of the no-loop configuration, $P_f(\Lambda_{h_N})$, has a scaling relation with the area of the lattice, and we have proposed the functional form (\ref{eq:Pfapprox}) for a general size of the honeycomb lattice as an approximation. We briefly report here that the probability of one-loop length-six configurations and that of length-ten can also be explained by the proposed form of $P_f$. This might suggest that the form of $P_f$ with other additional criteria can explain how the distribution of the total length of the loops takes such a peculiar form in Fig.\ref{fig:looplen_dist}. The area laws of the average loop length $\bar l$ and the average loop number $\bar \alpha$ are also to be explained. We do not know the answers yet. 
In the numerical calculations of the flow model (II), we have shown the probability of how many fluctuations of $\delta\hat G$ had been added in a million trials. As $N$ goes over $50$, the system seems to be stuck against the projected addition. This reduction of the fluctuations is entirely due to the definition of the projected addition. 
Because of the second line of the projection rule (\ref{def:addition}), if $G_1$ is huge enough as a set, many elements of $\hat {\cal L}(\Lambda)$ become non-additive. Even so, practically speaking in quasi-critical states, there are about 30\% 
vacant faces. 
Therefore, the system cannot be entirely frozen, changing only the time scale for its dynamics. 
It is the MSD on which the time scale matters. Especially, the emergence of the plateaus in Fig.\ref{fig:MSD11-41} might be because of this strong restriction of the fluctuations. We then expect that softening the projection in the addition may change the intermediate plateau part of the MSDs, and this point is to be explored more elsewhere \cite{IMTY}.

There are many related statistical models. Particularly, the vicious walk and the DLL model are the closest ones to the model (I). The vicious walk is the model of mutually avoiding multiple particle system so that its densest phase inevitably coincides with ours with the simplest stochastic process. The DLL model is the model of cooperative particle rearrangement for supercooled liquids and polymer melts, usually equipped with volume fluctuations and their temperature dependence. In both cases, the dynamics is governed by completely independent time steps, and they surely overlap with this sort of the models among the family of our models. Especially in the DLL model, the dynamics is usually equipped with the temperature dependence of $x(i)$, so it would be intriguing to investigate another manifestation of the temperature dependence and multi-component systems \cite{Polanowski} in both contexts. In addition, the loop gas model is one of the closest models. Letting all $P_f$ be one, one automatically obtains the partition function of the loop gas model from that of our stochastic Markovian model of the lattice liquid. 
The loop gas model can be described in the $O(N)$ spin model on the lattice by expressing all the loop graphs in terms of $O(N)$ spin variables \cite{Kostov89}. To avoid complications, we have not included this algebraic description of our models. Also, in the presence of the $P_f$ less than unity, the loop probabilities of both models become different. We can quantify them by the use of number series known as the derangement number and its related series. To invoke the expression, the models must be first written in the form of algebras. We have put aside this procedure, and therefore, to be given elsewhere \cite{IMTY}.

We are afraid that the constructed models are not so applicable and not ready to be compared to experiments, except for a few cases in a very different context. Because we used the honeycomb lattice as an explicit example, the current model might be able to be applied directly to a particle system of the same situation: a bigger sort of particles form a most densely packed crystal sitting at the centres of the faces while smaller particles moving in between them. Because of the set-up, the MSD of the flow model actually resembles that of the super-cooled polymer melt \cite{SuperCool}. Since their dimensions are different, we consider this happens by accident though.

In order to apply 2D lattice liquid models to soft-material membranes directly, one should remove or relax some of the assumptions we used in this paper for convenience. Namely, the assumptions are such that the system must be single-component, single-layered, incompressible, non-explicitly-viscous, less density fluctuating. Also, we have not used interactions between particles. Note that they are not the requirements nor the definitions of the 2D lattice liquid models. Among them, double-layer can easily be embedded just by folding a layer into two, but the interactions between two layers are not obvious. The simplest one is to introduce the interaction between layers only at the common boundary of two layers, possibly as a boundary condition. Our construction itself is ready to be extended to the multi-component case where there are more than one particle species and the interaction between different species should be introduced as well. The simplest one will be the Ising-type nearest neighbour interaction with which one can observe the phase separation according to some temperature. This might also turn on the effective viscosity in the model. Note the concept of the temperature can naturally be installed by letting $x,p$ be a function of $T$, for example, by $x\equiv e^{-\varepsilon/T}/deg$ and $p\equiv 1-x\cdot deg$. Thermodynamic behaviour will depend on the manifestation of the functional forms of $x$ and $p$. Some of the above proposals will be studied elsewhere \cite{IMTY}

Another very interesting yet possibly very hard model to solve is the fluid model with memory effect. As we mentioned in the manuscript, the ``potential'' probabilities can potentially carry the information from the past so that the past partially control the forthcoming event through the probabilities. Here, we have only dealt with Markov processes in both models, and the problem is beyond our scope of this manuscript.

\vspace*{10pt}
\noindent
{\bf Acknowledgements}

The authors acknowledge Prof. K. Yasuoka, Prof. Y. Masubuchi, Prof. S. Yasuda, Dr. T. Akimoto, and all the members of the group {\it Multiscale Simulation for Softmatters} for their stimulating discussions. One of the authors (YI) would like to thank Prof. K. Nagayama for his advice and encouragement, and S. Ishimoto for his support. This research was supported by {\it Core Research for Evolutional Science and Technology (CREST) of Japan Science and Technology Agency Corporation (JST)}.

\appendix
\section{On the space of the loop configurations}
\label{ap:loops}

To begin with, let us define the (oriented) loop configurations on $\Lambda$. For a given lattice $\Lambda$, a one-loop graph, or configuration, is defined by a sequence of sites and bonds with no repeating sites and no ends. The object is the same as simple cycles in graph theory. An oriented one-loop graph is defined in the same way and is given by a one-loop graph with an orientation of the bonds representing its sequence. The space of the non-oriented non-intersecting loop configurations ${\cal L}(\Lambda)$ on $\Lambda$ is the whole set of the non-intersecting combinations of such one-loop graphs. The space of the oriented non-intersecting loop configuration $\hat {\cal L}(\Lambda)$ is those of the oriented one-loop graphs. These spaces can be decomposed into (oriented) $k$-loop subspaces, or can be defined by them. 

Let ${\cal L}_i$ be a one-loop graph on $\Lambda$ for $i \in (1, \cdots, k)$. 
Then, a $k$-loop graph $G_k$ is defined by:
\bea
  G_k = \cup_{i=1}^{k} {\cal L}_i , {\rm ~s.t.~} {\cal L}_i \cap {\cal L}_j = \emptyset ~ {\rm ~for~} i \ne j.
\eea
Collecting all such $k$-loop graphs in ${\cal L}(\Lambda)$ constitutes the $k$-loop subspace ${\cal L}_k(\Lambda)$ of ${\cal L}(\Lambda)$:
\bea
  {\cal L}_k(\Lambda) \equiv \left\{ G_k \right\} , \quad 
  {\cal L}_k(\Lambda) \subseteq {\cal L}(\Lambda).
\eea
When $\Lambda$ is finite, there exists the maximum number of co-existing non-intersecting loops on $\Lambda$. Denoting it by $k_{max}$, one can write down the decomposition as:
\bea
&&
  {\cal L}(\Lambda) = \oplus_{k=0}^{k_{max}} {\cal L}_k (\Lambda), 
  \nn&&
  {\rm s.t.~} {\cal L}_i(\Lambda) \cap {\cal L}_j(\Lambda) = \emptyset ~ {\rm ~for~} i\ne j .
\eea
In the same manner, one can decompose the oriented space $\hat {\cal L}(\Lambda)$ into the oriented $k$-loop subspaces:
\bea
&&
  \hat {\cal L}_k(\Lambda) \equiv \left\{ \hat G_k \right\} , \quad 
  \hat {\cal L}_k(\Lambda) \subseteq \hat {\cal L}(\Lambda),
\nn&&
  \hat {\cal L}(\Lambda) = \oplus_{k=0}^{k_{max}} \hat {\cal L}_k (\Lambda), 
\nn&&
  {\rm s.t.~} \hat {\cal L}_i(\Lambda) \cap \hat {\cal L}_j(\Lambda) = \emptyset ~ {\rm ~for~} i\ne j ,
\eea
where $\hat G_k$ is an oriented $k$-loop graph in $\hat {\cal L}(\Lambda)$. 
If we introduce the function $\Omega$ which counts the number of elements of a set, then 
the uniqueness of the no-loop configuration in both spaces is trivially given by $\hat {\cal L}_0(\Lambda) = {\cal L}_0(\Lambda)$ and $\Omega({\cal L}_0(\Lambda)) = 1$. We label this unique configuration by $I(\Lambda)$. Similarly, $\Omega(\hat {\cal L}_1(\Lambda)) = 2 \,\Omega({\cal L}_1(\Lambda))$ due to the orientations. One can further observe the relation: $\Omega(\hat {\cal L}_k(\Lambda)) = 2^{k} \,\Omega({\cal L}_k(\Lambda))$ for $0<k\leq k_{max}$.

The algebraic structure of the space $\hat {\cal L}(\Lambda)$ directly corresponds to the physical relations between the states on $\Lambda$.
However, even for a very basic question of how many configurations are in the space $\hat {\cal L}(\Lambda)$, we have not known any clear answer yet for an arbitrary size of the lattice. In fact, it is shown that counting the number of self-avoiding walks (SAWs) for fixed ends on a given graph is $\#P$-complete \cite{SAWNP}. $\#P$ is a class of the problems for counting and, roughly speaking, they are expected to be as hard as $NP$-complete problems such as the prime number decomposition of an arbitrary large number. Therefore, one can expect that counting the number of the elements of $\hat {\cal L}(\Lambda)$ is as hard as $NP$-complete problems as well, because a non-intersecting loop is nothing but a self-avoiding walk ending at its start \cite{HCNP}. 

Below, we illustrate the structure of the spaces in terms of the cycle space in graph theory. 
We restrict ourselves to the planar cases unless otherwise stated. 
When $\Lambda$ is planar, {\it i.e.}, $\Lambda$ can be embedded on a plane with no bonds crossing, the space ${\cal L}(\Lambda)$ can be spanned by the set of the smallest loops which are the perimeters of the interior faces of $\Lambda$. In graph theory, there is the so-called cycle space $cycle(\Lambda)$ which is generated by all the simple cycles and whose basis is given by the fundamental cycles. 
The addition generates the cycle space from the basis, and makes it the vector space of $\Z_2^{E-V+1}$, where $E$ and $V$ are the number of the bonds and the number of the sites, respectively.
The concepts of ${\cal L}(\Lambda)$ and the cycle space $cycle(\Lambda)$ are very close to each other, but not always the same because of the presence of the intersecting loops in the cycle space. In fact, there is a necessary and sufficient condition for their equivalence that the degrees of the sites must be less than four everywhere: $deg(i)<4$ for all $i\in {\cal L}(\Lambda)$. 
Otherwise, the addition allows intersecting loops. Therefore, 
\bea
{\cal L}(\Lambda) \subseteq cycle(\Lambda),
\eea
in general, and ${\cal L}(\Lambda)$ is not necessarily a vector space but can be spanned by the set of the smallest loops. 

When $\Lambda$ is a planar honeycomb lattice of a finite size, 
\bea
{\cal L}(\Lambda)=cycle(\Lambda) \simeq \Z_2^{E-V+1} , 
\eea 
and
\bea
   \Omega({\cal L}(\Lambda)) = 2^{E-V+1} = 2^{F},
\eea
where $F$ denotes the number of the interior faces and the second equality is given by Euler's formula on a planar graph.
Its basis can be given by the set of the smallest loops.
Note that, in general, 
\bea
   \Omega({\cal L}(\Lambda)) \leq 2^{E-V+1}.
\eea
In the case of the oriented loop configurations $\hat {\cal L}(\Lambda)$, the relation is not as simple as above.
We can only provide an estimate for the number of the elements to close.
In the case of $k_{max}=\frac{F}{3}$, 
\bea
&&
    2^{F+1} -1 < \Omega(\hat {\cal L}(\Lambda_{h_{n,3m}})) <
    \nn&&\quad
    2^{\frac{4}{3} F } - 2^{\frac{F}{3}}(F+1) +2F + 1 .
\eea
A more refined relation will be given elsewhere \cite{IMTY}.

\section{The definition and the basic properties of the probability functions $P_f$ and $P_l$}
\label{ap:PfPl}

In this section, we define the probability functions of the no-loop configuration and of having (oriented) loops on the given graph, $P_f$ and $P_l$, respectively. That is to say, the complementary functions satisfying: $P_f + P_l = 1$. Upon their definitions, we will show some of the basic properties of the probability functions.
Note that the word `loop graph' indicates the same object as the loop configuration in the original context, but the latter emphasises an outcome of some stochastic process. 

On the loop graphs as events of the stochastic process, we define the following probability functions.
Let $P(G_k)$ be the probability of having the loop graph $G_k$ in $\Lambda$. Let $P_l(G)$ be the probability of having any loop graph on the graph $G^{\epsilon} \in \Lambda$. 
Let $P_f(G)$ be the probability of the graph $G^\epsilon\in \Lambda$ being completely frustrated, {\it i.e.}, having no loops. Here, the graph $G^\epsilon$ is an extended graph of $G$ such that it additionally includes all the bonds adjacent to the sites in $G$, in order to precisely explain and estimate the stochastic nature of our model.
Introducing the complementary graph $\bar {\cal L} \equiv \Lambda -{\cal L}$ to a loop graph ${\cal L}$, the probability function $P({\cal L})$ of having the loop graph ${\cal L}$ in $\Lambda$ is given by the product of the probability $P_l({\cal L})$ of the loop configuration ${\cal L}$ on the graph ${\cal L}$, and the probability $P_f(\bar{\cal L})$ of all the other sites being frustrated:
\bea
  P({\cal L}) 
        = P_l({\cal L}) \cdot P_f(\bar {\cal L}).
\eea
By definition, the following equalities and inequalities hold:
\bea
\label{eq: ineq}
  P_l(\Lambda) &=& \sum_{G_{k>0} \in {\cal L}(\Lambda)} P(G_k), 
  \nn
  P_f(G) &=& 1 - P_l(G), \quad P_f(\emptyset) = 1- P_l(\emptyset) = 1,
  \nn
  0 &\leq& P_l(G) \leq 1, \quad 0\leq P_f(G) \leq 1.
\eea
In the third line, 
for finite $G$ and $\Lambda$, the equalities $P_l(G)=1$ and $P_f(G)=0$ only hold when $G$ or $\Lambda$ is linear or the assignment of the ``potential'' probabilities of the particles' moving are unnaturally given in the non-stochastic limit.
Therefore, we assume $0\leq P_l(G)<1$ and $0<P_f(G)\leq 1$ for finite size $G$ and $\Lambda$.

Letting ${\cal L}$ be the naive projection of the oriented loop graph $\hat {\cal L}$ by removing the orientations, and assuming that the different oriented loop graphs which share the same projected loop graph ${\cal L}$ are equiprobable, then the mapping of the oriented loop graphs by the probability functions become apparent as:
\bea
   P(\hat G_k) &=& P_l(\hat G_k) \cdot P_f(\bar{G_k}),
   \nn
   P(G_k) &=& 2^k P(\hat G_k),
   \nn
   P_l(G_k) &=& 2^k P(\hat G_k),
   \nn
   P_f(\hat G_k) &=& P_f(G_k).
\eea 
$\hat G_k$ is not a subgraph of $\Lambda$ so that the complementary graph is defined by the projected graph $G_k$: ${\bar {\hat G}_k} = \Lambda - G_k = \bar G_k$. One can interpret that the same projection is done for the $P_f$ as well.

Because the frustrations in disconnected graphs are independent, the following relation holds: for two disconnected graphs, $G_1, G_2 \in \Lambda$:
\bea
  P_f(G_1 \cap G_2) = P_f(G_1) \cdot P_f(G_2) \quad {\rm for~} G_1 \cap G_2 = \emptyset.
\eea
Note that, using the above definitions and successive application of the relations, one can obtain $P(G_k)$ in a nested expression of the loop probabilities only.

\section{Asymptotic value of the mean square displacements (MSD)}
\label{app:aymptotic}
\subsection{Set-up and derivation of the exact asymptotic values of MSD}

In this section, we calculate the asymptotic value of the mean square displacements, $\braket{R^2(t)}$, which is the average of ${\braket{x(t)^2}}$ over $I$, in the limit of $t\to\infty$.
Before going into the actual calculation, we fix the notations and introduce some functions to facilitate it.

Number the position on $\Lambda$ starting from 0 to $M-1$, and assign a pair of integers to the $p$-th site, ${\mathbf p}=(p_x,p_y)$. 
The position of the $p$-th site, $(x_p,y_p)$, on the two-dimensional ${\R}^2$ plane can be given by, for the $N \times N$ honeycomb lattice:
\bea
&& p_x = p~ mod~ col, \quad  p_y = [p/col], \nn
&& x_p = \frac{\sqrt{3}}{2} p_x , \quad y_q = \frac32 p_y + \frac14 (-1)^{p_x+p_y+1} , 
\eea
where $M = dim(V(\Lambda)) = (N+1)(2N+1) = 2N^2 + 3N + 1, col = 2 N+1$, and $[A]$ denotes the greatest integer which does not exceed $A$. $p_x$ takes the value from $0$ to $(col-1)=2N$, while $p_y=0,\ldots, N$.
$p_{x,y}$ are all integers. The lattice spacing is taken to be a unit length.
The square distance between $p$-th and $q$-th positions are: 
\bea
&&
 R_{{\mathbf q}, {\mathbf p}-{\mathbf q}}^2
 = ( x_p - x_q )^2 + ( y_p - y_q )^2
   \nn&&
 = \frac34 \left\{ (p_x - q_x)^2 + 3 (p_y - q_y)^2 \right\}
 \nn&&
   + \frac18 \left\{1 - 6(-1)^{p_x+p_y} (p_y-q_y) \right\} 
   \left( 1 - (-1)^{(p_x-q_x) + (p_y-q_y)} \right).
   \nn
\eea
Note that we used the following identity: 
$1 = (-1)^{2k}$ for $\forall k\in\Z$.
Replacing ${\mathbf p}-{\mathbf q}$ by $(k,l)$, one can rewrite this as $R_{{\mathbf q},(k,l)}$: 
\bea
  \label{C3}
  R_{{\mathbf q},(k,l)}^2 
  &=& \frac34 k^2 + \frac94 l^2
   + \frac18 C({\mathbf q},l)
   - \frac{1}8 C({\mathbf q},l) (-1)^{k+l}, 
   \nn
\eea
where 
$
C({\mathbf q},l) \equiv \left( 1 + 6 l (-1)^{q_x+q_y} \right).
$
From this expression, one can verify that $R^2_{{\mathbf q},(k,l)}$ is integer.

Assume that, in the limit of $t\to +\infty$, all the pairs of the sites are independent to each other so that the asymptotic value of $\braket{R^2(t)}$ can simply be given by the average of the square distances of all the relevant pairs in $M\times M$ combinations. Namely, the value leads to the expression:
\bea
\label{eq:R_inf}
  \label{eq:R^2}
&&
  \braket{R_\infty^2} \equiv \lim_{t\to \infty} \braket{R^2(t)}
\nn&&= 
  \frac{1}{(M^2)_{relevant}} \left[ 
  \sum_{q_y=0}^{N} \sum_{q_x=0}^{col-1} \sum_{l=-q_y}^{N-q_y} 
  \sum_{k=-q_x}^{col-1-q_x} 
  R^2_{q,(k,l)}
  \right]_{relevant} \!\!\!\!\!\!\!\!\!\! . \nn
\eea
Here, the word ``relevant'' means that some sites at the boundary cannot contribute to the above, because our flow dynamics can only involve those points which have more than one bond. For instance, regardless of the value of $N$, the following two sites should always be excluded:
\bea
  q = \left\{ (0,0), (2N,0) \right\}.
\eea
Besides, when $N$ is odd, the following two sites are to be taken into account in the same manner:
\bea
  q = \left\{ (0,N), (2N,N) \right\}.
\eea
They are connected only one nearest neighbouring site, so that the particle cannot reach these four points via our flow dynamics. Therefore, the square distances related to these points do not contribute to the above calculation. Substituting (\ref{C3}) into (\ref{eq:R^2}) and subtracting such unnecessary terms from the sum, we obtain for odd $N$: 
\bea
&&
  \braket{R^2_{\infty}}
  \nn&&
  = \frac{1}{\left\{ (N+1)(2N+1)-4+(1+(-1)^N) \right\}^2} \times
  \nn&&
  \Biggl[
    \frac18 (N+1)^2 (2N+1)^2 ( 7N^2+10N+1 ) 
    \nn&&\quad
    + \frac34 (N+1)^2 (2N+1) 
    \nn&&\quad
    - 2\left( 14N^4 + 29N^3 - N^2 +N +1  \right)
    \nn&&\quad
    -\frac18 \left\{ 6(N+1)^2(2N+1) + 1 \right\} 
      \left( \frac{1+(-1)^{N}}2 \right) 
    \nn&&\quad
    + \left(  
    14N^4 + 29N^3 - 16N^2 - 2N + 2
      \right) \left( \frac{1+(-1)^{N}}{2} \right)
  \Biggr] 
\nn&&
  \simeq \frac18 (7 N^2 + 10 N ) ,
\eea
up to $O(1)$. 
In the cases of $N=11, 21, 31, 41, 51$, the exact values are given in the manuscript up to first five digits.



\begin{thebibliography}{29}
\expandafter\ifx\csname natexlab\endcsname\relax\def\natexlab#1{#1}\fi
\expandafter\ifx\csname bibnamefont\endcsname\relax
  \def\bibnamefont#1{#1}\fi
\expandafter\ifx\csname bibfnamefont\endcsname\relax
  \def\bibfnamefont#1{#1}\fi
\expandafter\ifx\csname citenamefont\endcsname\relax
  \def\citenamefont#1{#1}\fi
\expandafter\ifx\csname url\endcsname\relax
  \def\url#1{\texttt{#1}}\fi
\expandafter\ifx\csname urlprefix\endcsname\relax\def\urlprefix{URL }\fi
\providecommand{\bibinfo}[2]{#2}
\providecommand{\eprint}[2][]{\url{#2}}


\bibitem[{\citenamefont{Sakuma}(2010)}]{SakumaIT}
\bibinfo{author}{\bibfnamefont{Y.} \bibnamefont{Sakuma}},
  \bibinfo{author}{\bibfnamefont{T.} \bibnamefont{Taniguchi}}, 
  \bibnamefont{and}
  \bibinfo{author}{\bibfnamefont{M.} \bibnamefont{Imai}},
  \bibinfo{journal}{Biophys. J. {\bf}} \textbf{\bibinfo{volume}{99}},
  \bibinfo{pages}{472-479} (\bibinfo{year}{2010}).

\bibitem[{\citenamefont{Lipid Bilayer}(zzzz)}]{Lipids}
\bibinfo{author}{\bibfnamefont{A. D.} \bibnamefont{Bangham}},
  \bibinfo{author}{\bibfnamefont{M. M.} \bibnamefont{Standish}},
  \bibnamefont{and}
  \bibinfo{author}{\bibfnamefont{J. C.} \bibnamefont{Watkins}},
  \bibinfo{journal}{
  J. Mol. Biol.} \textbf{\bibinfo{volume}{13}},
  \bibinfo{pages}{238-252} (\bibinfo{year}{1965}).

\bibitem{LipidSim}
H. Heller, M. Schaefer, and K. Schulten, 
J. Phys. Chem., {\bf 97} (1993) 8343-8360; 
D. P Tieleman, S. J. Marrink,H. J. C. Berendsen, 
Biochimica et Biophysica Acta {\bf 1331} (1997) 235-270.


\bibitem{Surfactant}
O. Holderer, M. Klostermann, M. Monkenbusch, R. Schweins, P. Lindner, R. Strey, D. Richter and T. Sottmann, 
Phys. Chem. Chem. Phys. {\bf 13} (2011) 3022-3025.

\bibitem{Various}
R. Lipowsky, 
Nature {\bf 349} (1991) 475-481; 
E. Sackmann, 
FEBS Lett. {\bf 346} (1994) 3-16.

\bibitem{Simulation}
H. Noguchi and G. Gompper, 
Phys. Rev. E {\bf 73}, 021903 (2006); 
T. Akimoto, E. Yamamoto, K. Yasuoka, Y. Hirano, and M. Yasui, 
Phys. Rev. Lett., {\bf 107}, 178103 (2011);  
H. Noguchi, 
J. Chem. Phys. {\bf 134}, 055101 (2011); 


\bibitem{FieldTheory}
K. Katsov, M. Muller, and M. Schick, 
Biophys. J. {\bf 87} (2004) 3277-3290.

\bibitem[{\citenamefont{Fisher}(1984)}]{fisher}
\bibinfo{author}{\bibfnamefont{M. E.}~\bibnamefont{Fisher}}, 
  \bibinfo{journal}{
  J. Stat. Phys. {\bf 34}} (\bibinfo{year}{1984}) 667-729.

\bibitem[{\citenamefont{Pakula}(1997)}]{Pakula}
\bibinfo{author}{\bibfnamefont{T.} \bibnamefont{Pakula}} \bibnamefont{and}
  \bibinfo{author}{\bibfnamefont{J.} \bibnamefont{Teichmann}},
  \bibinfo{journal}{Mater. Res. Soc. Symp. Proc.} \textbf{\bibinfo{volume}{455}}
  (\bibinfo{year}{1997}) \bibinfo{pages}{211-222}; 
\bibinfo{author}{\bibfnamefont{T.} \bibnamefont{Pakula}},
  \bibinfo{journal}{J. Mol. Liq.} \textbf{\bibinfo{volume}{86}}
  (\bibinfo{year}{2000}) \bibinfo{pages}{109-121}.
 
\bibitem[{\citenamefont{Kostov}(1989)}]{Kostov89}
\bibinfo{author}{\bibfnamefont{B.} \bibnamefont{Nienhuis}},
  \bibinfo{journal}{Phys. Rev. Lett.} \textbf{\bibinfo{volume}{49}}
  (\bibinfo{year}{1982}) \bibinfo{pages}{1062-1065};
\bibinfo{author}{\bibfnamefont{F.} \bibnamefont{Di Francesco}},
  \bibinfo{author}{\bibfnamefont{H.} \bibnamefont{Saleur}},
 \bibnamefont{and}
  \bibinfo{author}{\bibfnamefont{J.-B.} \bibnamefont{Zuber}},
  \bibinfo{journal}{Nucl. Phys. {\bf B}} \textbf{\bibinfo{volume}{285}}
  (\bibinfo{year}{1987}) \bibinfo{pages}{454-480}; 
\bibinfo{author}{\bibfnamefont{I.~K.} \bibnamefont{Kostov}},
  \bibinfo{journal}{Nucl. Phys. {\bf B}} \textbf{\bibinfo{volume}{326}}
  (\bibinfo{year}{1989}) \bibinfo{pages}{583-612}.

\bibitem[{\citenamefont{Ishimoto, Murashima, Taniguchi, and Yamamoto}(2012)}]{IMTY}
\bibinfo{author}{\bibfnamefont{Y.}~\bibnamefont{Ishimoto}}, 
\bibinfo{author}{\bibfnamefont{T.}~\bibnamefont{Murashima}}, 
\bibinfo{author}{\bibfnamefont{T.}~\bibnamefont{Taniguchi}}, 
  \bibnamefont{and}
  \bibinfo{author}{\bibfnamefont{R.}~\bibnamefont{Yamamoto}},
  \bibinfo{journal}{in preparation}  (\bibinfo{year}{2012}).



\bibitem[{\citenamefont{Polanowski}(2003)}]{Polanowski}
\bibinfo{author}{\bibfnamefont{P.} \bibnamefont{Polanowski}} \bibnamefont{and}
  \bibinfo{author}{\bibfnamefont{T.} \bibnamefont{Pakula}},
  \bibinfo{journal}{J. Chem. Phys.} \textbf{\bibinfo{volume}{118}}
  (\bibinfo{year}{2003}) \bibinfo{pages}{11139}. 


\bibitem[{\citenamefont{Binder}(20)}]{SuperCool}
J. Baschnagel, C. Bennemann, W. Paul and K. Binder, 
J. Phys.: Condens. Matter {\bf 12} (2000) 6365-6374.


\bibitem[{\citenamefont{Valiant}(1979)}]{SAWNP}
\bibinfo{author}{\bibfnamefont{L. G.}~\bibnamefont{Valiant}}, 
  \bibinfo{journal}{
  Theor. Comput. Sci. {\bf 8} (1979) 189-201. 
  }  \bibinfo{year}{}


\bibitem[{\citenamefont{Garey}(1976)}]{HCNP}
\bibinfo{author}{\bibfnamefont{M.}~\bibnamefont{Garey}}, 
  \bibnamefont{}
  \bibinfo{author}{\bibfnamefont{et}~\bibnamefont{al.}},
  \bibinfo{journal}{
  SIAM J. Comput. {\bf 5}}  (\bibinfo{year}{1976}) 704-714.


\end{thebibliography}
\end{document}